\documentclass[letterpaper,english,letterpaper,english,aps,pre,onecolumn,nobibnotes,nofootinbib]{revtex4}
\usepackage{mathptmx}
\usepackage[T1]{fontenc}
\usepackage[latin9]{inputenc}
\usepackage{units}
\usepackage{graphicx}

\makeatletter

\newcommand{\binom}[2]{{#1 \choose #2}}

\@ifundefined{textcolor}{}
{%
 \definecolor{BLACK}{gray}{0}
 \definecolor{WHITE}{gray}{1}
 \definecolor{RED}{rgb}{1,0,0}
 \definecolor{GREEN}{rgb}{0,1,0}
 \definecolor{BLUE}{rgb}{0,0,1}
 \definecolor{CYAN}{cmyk}{1,0,0,0}
 \definecolor{MAGENTA}{cmyk}{0,1,0,0}
 \definecolor{YELLOW}{cmyk}{0,0,1,0}
}

\makeatother

\usepackage{babel}
\begin{document}

\title{Expertise localization discovered through correlation of key term distribution and community detection in co-author networks}

\author{Joe Durante}

\affiliation{Booz Allen Hamilton Inc., 500 Avery Lane, Suite C, Rome, NY 13441}

\affiliation{Cornell University, 410 Thurston Avenue, Ithaca, NY 14850}

\author{Tyler Whitehouse}

\affiliation{Booz Allen Hamilton Inc., 3811 N. Fairfax Dr., Ste. 600, Arlington,
Virginia 22203}

\author{Gino Serpa}

\email{serpa_gino@bah.com}

\affiliation{Booz Allen Hamilton Inc., 3811 N. Fairfax Dr., Ste. 600, Arlington,
Virginia 22203}

\author{Artjay Javier}

\affiliation{Booz Allen Hamilton Inc., 3811 N. Fairfax Dr., Ste. 600, Arlington,
Virginia 22203}

\begin{abstract}
We present an efficient and effective automatic method for determining
the research focus of scientific communities found in co-authorship
networks. It utilizes bibliographic data from a database to form the
network, followed by fastgreedy community detection to identify communities
within large connected components of the network. Text analysis techniques
are used to identify community-specific significant terms which represent
the topic of the community. In order to greatly reduce computation
time, the `Topics' field of each publication
in the network is analyzed rather than its entire text. Using this
text analysis approach requires a certain level of statistical confidence,
therefore analyzing very small communities is not effective with this
technique. We find a minimum community size threshold of 8 coauthored
papers; below this value, the community's topic cannot be reliably
identified with this method. Additionally, we consider the implications
this study has regarding factors involved in the formation of scientific
communities in co-authorship networks.
\end{abstract}
\maketitle

\section{Introduction}

Network analysis has provided insights on a wide variety of subjects from unraveling the leadership structure in a terrorist network \cite{Hey} to understanding the spread of computer viruses \cite{PhysRevLett.86.3200}.
Networks are present in many aspects of science, technology and even
everyday life, such as social networks that describe different types
of interactions between people. Here we study a particular type of
social network: co-authorship networks. These networks capture interactions
between researchers as connections are formed between two or more
authors when they collaborate to publish a scientific paper. Co-authorship
networks are of interest because if one can determine how the field
is evolving \cite{Barabasi2002590,albert2002statistical}, one can
make informed predictions about the direction of the field and where
advancements are most likely to occur.

\begin{figure}
\includegraphics[scale=0.4]{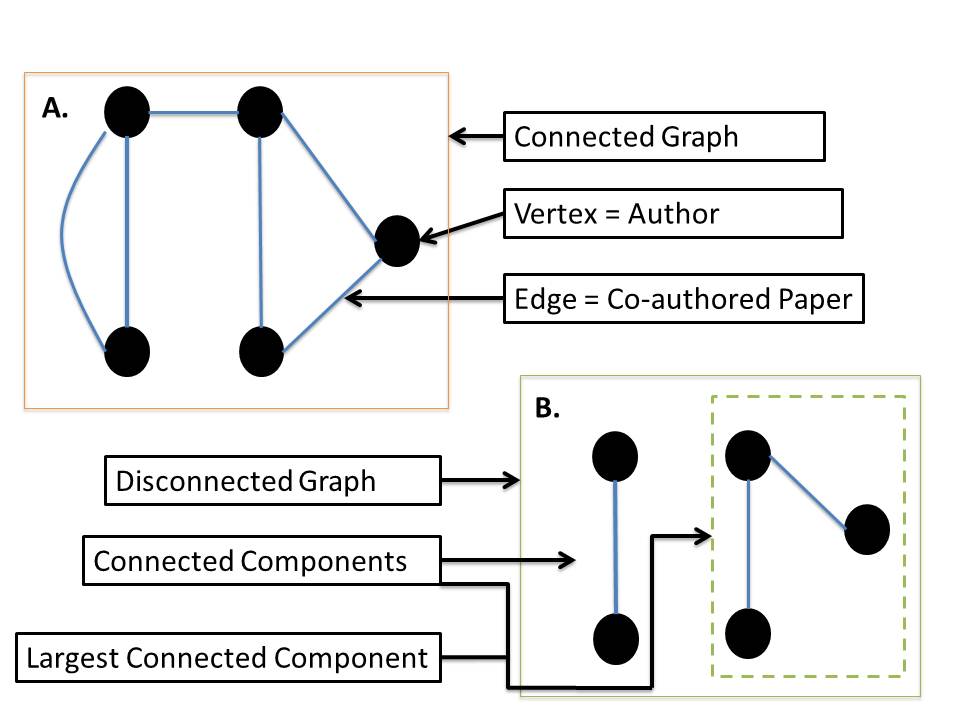}\protect\caption{Basics of graph representation and co-authorship network. This is a simple pictorial example of how a co-authorship
network is represented by a graph. Vertices (dots) represent authors in the network while an edge (line) represents a paper
that was co-authored by the authors it connects. Whereas each vertex
is a unique author, an edge is not necessarily a unique paper. For
example, the graph in A has 5 vertices and 6 edges. Therefore, the
network represented by the graph definitely consists of 5 authors.
However, based on this specific graph structure we cannot be certain
if there are 4 or 6 papers. This is because the triangle formed on
the right hand side can be formed by three papers, each written by
two authors; or it could represent one paper written by all three
authors where each edge represents the same paper. The example in A is a connected
graph, while B is a disconnected graph. The network in B has 2 components
where the component in the dashed box is its Largest Connected Component based
on the fact it contains the most authors.\label{fig:Network-Basics} }
\end{figure}

Networks are commonly represented by graphs. An example is seen in
Figure \ref{fig:Network-Basics} which shows a simplified co-authorship
network. Here the individual vertices represent authors and the co-authored
publications represent edges that connect the vertices. If two or
more authors publish a paper together, then they are connected in
the graph by the edge representing that paper. When a graph is formed,
there may or may not be a path between all pairs of vertices. If there
is a connecting path between all vertices in a graph, the graph is
connected as seen in Figure \ref{fig:Network-Basics}A. Otherwise,
the graph is disconnected as in Figure \ref{fig:Network-Basics}B.
However, connected components (or simply components) exist in disconnected
graphs. These are any group of vertices with a connection between
all pairs of vertices. The graph in Figure \ref{fig:Network-Basics}B
has two components, and the component in the dashed box is its Largest
Connected Component (LCC). Co-authorship networks, when formed from
enough authors and publications, will most often be disconnected and
exhibit connected components of varying sizes.

\begin{figure}
\includegraphics[scale=0.4]{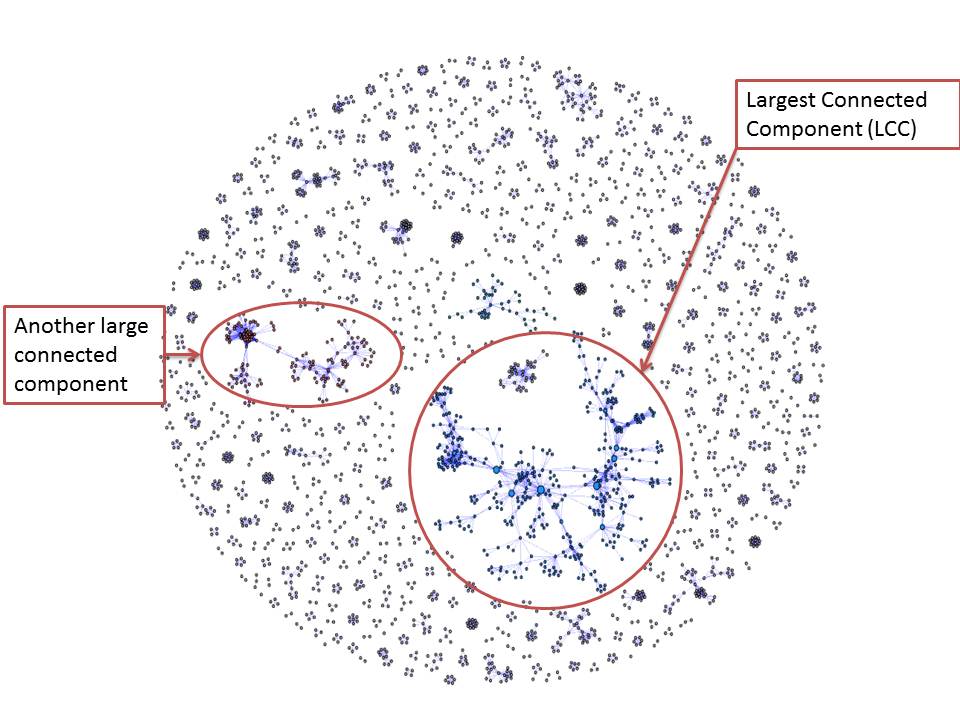}

\protect\caption{Graph representation of one of the co-authorship networks analyzed in this study.  On the outer level,
grouping first takes place as the network divides into many components.
The network has a few large components consisting of many authors;
the two largest are circled.  Additionally, there are
numerous small components, each formed by relatively few authors.
In small components, relationships between authors are clear and can
be easily identified with graph analysis tools. In large components,
tightly bound groups known as communities form. This is the inner
level of clustering. The communities are only loosely connected to
other communities in the component. Thus, to identify which authors
are collaborating closely, one must perform community detection on
larger components.\label{fig:A-co-authorship-network}}
\end{figure}

\begin{figure}
\includegraphics[scale=0.4]{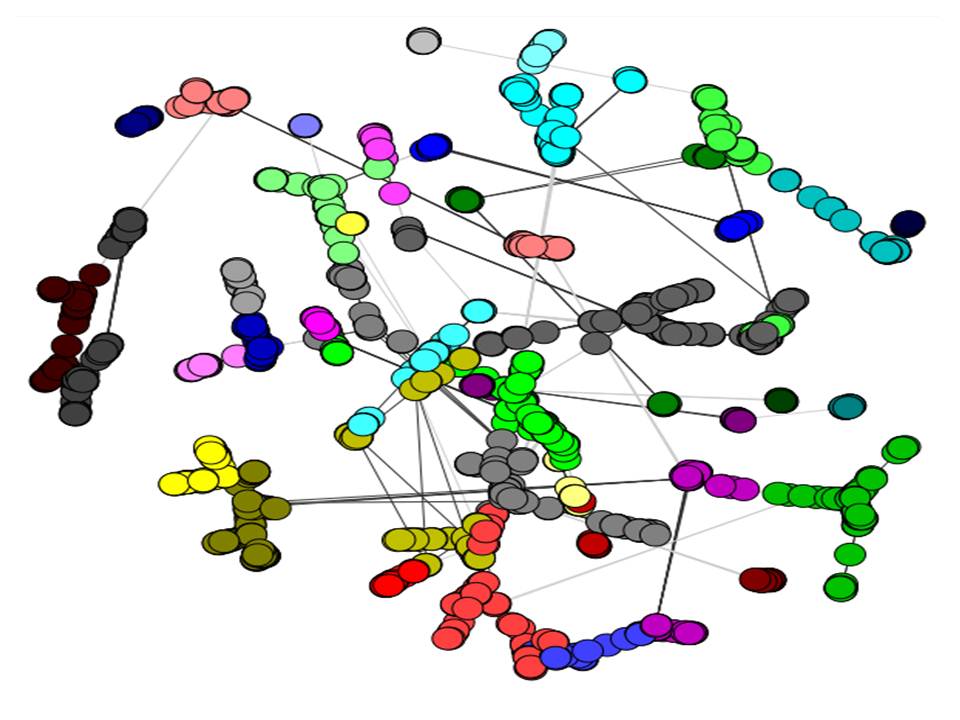}\protect\caption{Community Detection. Using Newman's method
of fastgreedy community detection, communities were identified within
a large component similar to the LCC circled in Figure \ref{fig:A-co-authorship-network}.
This is the LCC of one of the networks analyzed in this study. Each different
color represents a community. Notice the clustering
present within the component. Though difficult to see, each community
is tightly bound by a large number of co-authored papers within the
group. However, the looser connections between the communities is
quite clear as there are relatively few edges between groups.\label{fig:Community-Detection}}
\end{figure}

Co-authorship graphs are used to determine which scientists in the
field are collaborating together. Then, based on available information
about those scientists (such as papers they have written), one may
infer what type of research is taking place in that group and where
it is happening. In small components the relationships between authors
are relatively easy to see and pick out using graph analysis tools.
However, co-authorship networks, like those in Figure \ref{fig:A-co-authorship-network},
can grow vast connected components with thousands of authors \cite{Newman2001_CoAuthorNet,Barabasi2002590}.
Though all the authors share common connections in a connected component
of such magnitude, this does not imply that they are all actually
working together. Instead, a few prominent researchers connect many
different groups of researchers who in large part do not otherwise
collaborate with each other \cite{Girvan11062002,palla2005uncovering,Newman2001_CoAuthorNet}.
Co-authorship networks exhibit the following type of hierarchical
clustering. The entire network is comprised of connected components
of various sizes. This is the first level of groupings. Then, the
larger connected components are composed of several tightly bound
groups of authors loosely connected by a few co-authors common to
more than one group. These tightly bound groups are the second level
of clustering \cite{PhysRevE.69.066133,Newman2001_CoAuthorNet}. Figure
\ref{fig:A-co-authorship-network} illustrates a network with this
hierarchical clustering structure and Figure \ref{fig:Community-Detection}
looks at the clustering present within a large component more closely.
Gathering information about which authors are frequently collaborating
within large components often requires detecting these smaller, more
closely connected groups of authors, known as communities. There are
many algorithms for community detection, each of which has its advantages
and drawbacks. For example, oftentimes, some accuracy is sacrificed
for savings in computation time \cite{Fortunato02012007}. Here we
implement Newman and Girvan\textquoteright s method of greedily optimizing
modularity \cite{PhysRevE.69.026113}, known as fastgreedy. Figure
\ref{fig:Community-Detection} illustrates how communities were identified
by fastgreedy within one of the large components used in this study.
Each different color represents a community. Though it is difficult
to see, many edges link vertices within communities. On the other
hand, there are relatively few connections between communities. Thus,
this component exhibits the hierarchical clustering structure described
earlier.

Fastgreedy is well established and successful at picking out tightly
bound groups from a whole. For example, it has been proven that fastgreedy
can successfully infer the conference structure of NCAA Division I
football by analyzing a graph in which vertices represent teams and
edges represent games between teams \cite{PhysRevE.69.066133}. There
are plenty of other case studies confirming the success of fastgreedy.
Thus, we know that the results returned by this community detection
generally provide an accurate representation of the scientific
communities present in a graph. A question we pose is, do these communities
reflect groups of researchers with an even narrower research focus
than the component of which they are a part? In other words, does the hierarchical
clustering present in the structure of the network imply a hierarchy
on the specificity of research topic also? If this is true, consider,
for example, forming a network of all authors who write about `animals'.
One connected component would have authors who focus on `dogs', while
another group would be concerned with the study of `cats'. Within
the `dogs' group, we expect to see communities with authors more intently
focused on subtopics such as `golden retrievers' and `beagles'. Or,
on the other hand, let's assume that research topic specificity is not the case. Instead, only
regional factors and institutional affiliations determine the hierarchy
of the network. Then the United States would have its own component
with authors writing papers about various topics in the `animals'
category. Within that component would be a community consisting of
authors from the Cornell University College of Veterinary Medicine
also writing about all types of animals. As a result, the communities
would not represent groups of authors focused on a specific topic
but rather groups of authors sharing a location and affiliation. Likely
a combination of these factors, and others, determines the structure
of the network and how communities are composed. However, in this
study, we exclusively examine if the division of a component into communities
implies that the individual communities are focused on more specific
research goal topics than within the entire component.

In addition, this study also focuses on developing and validating an automatic way to determine
the topics of a community. Our method utilizes bibliographic data
from a database to form the network, community detection to pick out
scientific communities within the network, followed by text analysis
techniques to find the topics of those communities. By analyzing a
community's papers, these techniques extract important terms that
are unique to the community. When examined together, the terms should
provide a clear idea of the community's topics. This approach will
rely on a level of statistical confidence which requires that the
communities evaluated be large enough to be statistically significant. Thus, we propose a minimum
community size threshold- the community should have at least 8 papers
or 15 authors- necessary to produce meaningful results from the procedure.
If, through the use of this method, most communities are shown to
exhibit increased concentration on certain terms, this is an indicator
that the communities are indeed focused on specific research topics.

The remainder of the paper is divided into four sections. Section
II describes the theory behind the text analysis techniques used to
identify words that relate to the topic of a community. Section III
describes the automatic method of using bibliographic data to form
a network, followed by community detection, and the text analysis
techniques we use to find terms that represent the topics of the communities.
Section IV discusses the effectiveness of this method and suggests
a minimum community size necessary for using the method. Section V
summarizes the paper and discusses the implications this study has
regarding the factors behind formation of scientific communities in
co-authorship networks.

\section{Text Analysis Techniques}

How do we determine if communities identified by a detection method
represent groups of researchers publishing papers focused on a common
topic? A qualitative way would be to isolate a community, gather its
papers and then examine them for content. Of course, this is imprecise
as differing opinions will lead to different conclusions about the
cohesiveness of a community. Additionally, to do this manually for
every community in a component would be laborious and time consuming.
Here we present an automatic method to extract important terms
from the publications of a community in order to identify the topic of the community.

Consider all the documents (this can either refer to the papers,
or collections of words representing the content of the papers, such
as the abstracts of papers) written inside a community created not
by their authors, but by picking words one at a time at random from
a set of possible words. This is a process known as sampling with
replacement; it is like picking words from a "bag
of words" where, after a word is picked and marked
down, it is put back in the bag. For our experiment the universe of
available words will be all the words present in any document in the
component of which the community is a part. Continuing the example
from Section I, let's say the component focused on `dogs' contained
the words in the set $\Omega$=\{`dog', `golden retriever', `puppy',
`beagle'\}. Each word has a different probability of being chosen
from the set which is estimated from examining how often the word
appears in the community. Consider a community in which the probability
of each word in the set $\Omega$ being picked was $P{}_{comm}=\left\{ \frac{1}{3},\frac{1}{6},\frac{1}{3},\frac{1}{6}\right\} .$
Altogether the probabilities of every possible word in the sample
space form a Probability Mass Function (PMF) from which to create,
one word at a time, all the documents in a community. A document formed
with $P{}_{comm}$ might look like this: {[}beagle dog beagle puppy
golden retriever puppy puppy{]} where words don't necessarily appear
with frequency equal to their probabilities in $P{}_{comm}$ as the
creation of the document is viewed as a random process. We can repeat
this argument to conceive the formation of any document with an underlying
PMF inferred not from the community to which it belongs, but instead from
the entire component; call this PMF $P{}_{comp}$.

Assume now that the separation of the component into communities had
no implications about focus on certain topics, i.e. the communities
are properly separated, but their composition is determined by other
factors besides striving towards a highly specific research goal.
Then, a community would contain documents that have nothing more in
common with each other besides the broader theme which defines the
component. Thus, one would not expect any certain words to be significantly
more probable within a community than throughout the entire component.
We would expect to see the PMF of the community to look very similar
to that of the component. In our example, if we had a community with
$P{}_{comm}=\left\{ \frac{21}{64},\frac{17}{64},\frac{11}{64},\frac{15}{64}\right\} $
and a component with $P{}_{comp}=\left\{ \frac{1}{3},\frac{1}{4},\frac{1}{6},\frac{1}{4}\right\} $,
all the words in $\Omega$ appear with nearly equal probability in
the `dogs' component and in the examined community. Thus, we would
conclude that no terms are unique to the community. This implies the
community is not focused on a topic more specific than `dogs' and
the division of the component into a community does not imply a narrowing
of research focus.

On the other hand, consider picking out a community from the component
in which scientists in the community pursued a mutual and specific
research goal. For example, in a graph of authors who research and
publish papers on dogs, we may have a group who concentrates on golden
retrievers. We should expect to see words pertaining to that goal
have a much higher probability of appearing in that community than
in the component on the whole. In the example, we expect the probability
of seeing `golden retriever' in
a document to be much higher in the community than in the component,
while the probability of seeing `dog'
should be about the same in both. So now $P{}_{comm}$ and $P{}_{comp}$
might look like this: $P{}_{comm}=\left\{ \frac{22}{64},\frac{1}{2},\frac{6}{64},\frac{4}{64}\right\} ,\: P{}_{comp}=\left\{ \frac{1}{3},\frac{1}{4},\frac{1}{6},\frac{1}{4}\right\}$.
In such communities, the words found to have a significantly higher
probability of appearing in the community than in the component can
be identified and then be further analyzed to determine the topic(s)
of the community. Notice `golden retriever' is the only word where
$p{}_{comm}$ (lower case $p$ is used to denote the probability of
one term appearing while capital $P$ denotes the entire PMF) is much
greater than $p{}_{comp}$. Thus, in this very simple example, after
examining the PMF's of the community and the component, we would identify
`golden retriever' as a significant term and correctly conclude the
community is concentrated on golden retrievers.

\section{Methodology}

\subsection{Forming a Co-authorship Network and Identifying Communities}

\subsubsection{Extracting Data From the Inspec Database to Form a Graph}

We obtain the scientific publications necessary to form co-authorship
networks using Thompson Reuter's ISI Web of Knowledge v.5.5 index to
access the Institute for Engineering and Technology's Inspec publication
database. The Inspec database contains an expansive collection of
papers in the fields of mathematics, physics, and computer science.
Once the Inspec database was selected through the Web of Knowledge
search interface, the Boolean keyword or series of keywords best representing
the sub-field under investigation were entered into the Inspec search
field. The searches were carried out over the time span 1969-2012,
the longest available in the database, although most searches returned
a group of publications spanning a shorter period. A keyword search
usually returned anywhere from $10^{1}$ to $10^{5}$ papers. Using
a custom Python script, the data was pre-processed by sub-fields to
extract a list of authors. The authors' last names and initials were
stored, while repetitions were removed. Information,
such as papers written and affiliation, was stored for each author. The same was done
for each paper returned by a search; information such as the paper\textquoteright s
authors and the string of key terms used to identify the topic of
a paper were extracted and stored.

\subsubsection{Creating a Graph and Identifying Communities }

The data from the publications was used to create a co-authorship
network. As mentioned before, to begin forming a graph, each author
is considered to be a vertex and an edge between two vertices is formed
for every paper that the two researchers co-authored together. Again,
there is most often not a path connecting all vertices in the graph,
and we see several connected components. Though a graph may have a
few large connected components, per graph we will only focus on the
LCC. Graphs will exhibit hierarchical clustering as the larger connected
components contain several groups of strongly interconnected vertices
within them. To pick out these communities in the LCC, the fastgreedy
community detection method was used.

\subsection{Extracting Key Terms to Represent Documents, Forming PMF's with Estimates
of $p$ for Each Key Term}

\subsubsection{Extracting Key Terms to Represent the Content of the Papers}

Each paper provides key terms under a `Topics' field that is useful
for searching the paper in a database. These terms provide a very
succinct representation of the content of a publication and will be
our device for inferring topic focus, or lack thereof, within communities.
They will serve as the 'words' discussed in Section II, while a paper's
collection of key terms will form the `documents'. A paper will have
anywhere from 5 to 70 key terms with an average paper containing 15
to 25 terms. This is a valid substitute for analyzing the entire text
of a paper because the key terms summarize the most useful information
(from the viewpoint of this study) that could be drawn from a paper
by identifying the topics of a paper. Additionally, it is possible
to circumvent analyzing the occurrences of common words that would
appear with about equal frequency in almost any paper, such as `the'
and `a'. Furthermore, by avoiding analyzing all the text in every
paper, computation time is significantly reduced by only performing
the text analysis techniques on the papers' key terms%
\footnote{Within each paper, a key term will not be repeated. Thus, if the documents
are viewed as being formed by picking words from a set, the formation
of each separate document should be viewed as sampling without replacement
instead of sampling with replacement. However, the `Topics' fields
are generally fairly short, containing about 20 key terms on average,
and $p{}_{comm}$ will be generally be much smaller than $\nicefrac{1}{20}$
for all terms. Thus, even if the documents were sampled with replacement
it is unlikely that a word would be picked twice. Additionally, documents
in both the LCC and the communities are being approximated as being
sampled with replacement from PMF's so any errors in the study originating
from this slight approximation of document formation will be negligible.%
}.

A custom Python script was written to identify each key term present
in the `Topics' fields of a paper. We
used this to identify all unique key terms in the LCC of a graph.
Several key terms had the same meaning, but were identified as distinct
terms due to spelling or phrasing differences. For example, `dog'
and `dogs' would be distinct terms. However, since their meaning is
the same, and this method is concerned with identifying topics, it
would make sense if they were viewed as the same term. For this reason,
groups of key terms with slight spelling or phrasing differences were
identified and one of the key terms was used to represent the entire
group. Whenever any term in the group appeared in a paper's `Topics'
field it was replaced by the group's representative key term. Once
these meaningless spelling or phrasing differences were accounted
for, we repeated the process of finding all unique key terms in the
LCC to form our set of words, $\Omega$.

\subsubsection{Estimating $p$, the Probability of a Word Being Chosen to Form a
Document}

Taking the view discussed in Section II in our analysis, that documents
are formed by sampling words from a set using the PMF of the community
to which it belongs, we first excluded papers that overlap into two or more
communities to avoid situations where a paper is formed by more than
one PMF. Fortunately this is a small percentage of the total papers
in a component as communities will be linked by a few papers, but
a large majority of the papers are contained entirely within the individual
communities. 

Then, for each key term in $\Omega$, a count was kept of the number of papers in which that term appeared within each community and also the total number that it appeared in the entire LCC. Let this count equal $k$; $k_{comm}$ for the count within
a community, $k{}_{LCC}$ for the count within the LCC. Additionally
the total count of all key terms was recorded for both the communities
and the LCC. This is not counting unique key terms, but rather it
is a sum of the number of key terms in all the papers in a community
or the LCC. Let this total count equal $n$; again $n{}_{comm}$ for
within a community, and $n{}_{LCC}$ for the LCC. Another way to think
of $n$ is that it is the sum of all the counts within the group of
interest for all terms belonging to $\Omega,$ so $n_{comm}=\sum_{\left\{ terms\in\Omega\right\} }k_{comm}$ and $n{}_{LCC}=\sum_{\left\{ terms\in\Omega\right\} }k_{LCC}$. In
each case, $k$ can be thought of as an observation of $K$, a binomially
distributed random variable with parameters $n$ and $p$. Here, $p$ is
what we are seeking to form a PMF for the community/LCC. However $p$
is unknown to us as we only observe $k$ and $n$. Thus we needed
an estimate, $\hat{p}$, of $p$ based on $n$ and our sole observation
of $K$. A simple estimate of $p$ would be the point estimate $\hat{p}=\frac{k}{n}$;
this is how the normal approximation to the binomial estimates $p$.
However, when $p\times n\leq5$, $\hat{p}=\frac{k}{n}$ is proven
to be an unreliable estimate of $p$ \cite{brown2001}. Since key
terms typically exhibited very low counts, $k$, compared to total
words, $n$, we assume it is likely that the true probability, $p$,
was very small also. Then, for many key terms $p\times n\leq5$, and
using $\hat{p}=\frac{k}{n}$ would not be a reliable estimate. Thus,
to test the appropriateness of other methods for estimating $p$ over
ranges of possible $k$ and $n$ seen in this study, the following
experiment was performed.

Given a term has true probability $p$ of being picked from a set,
consider a trial in which choosing that term is a success and picking
any other term is a failure. This describes a Bernoulli trial with
probability $p$ of success and can be represented by flipping a biased
coin with the probability of heads appearing equal to $p$. Repeating
this trial $n$ times will result in a binomial distribution for the
random variable $K$ representing the number of heads that appear
in $n$ flips of the coin where the probability of observing $k$
heads in the $n$ flips is: $P(K=k)=\binom{n}{k}(p)^{k}(1-p)^{n-k}$
for $k=0,1,2,...,n$. Using a $p$ commonly seen in key terms within
a community, we perform $n$ trials, observe $k$ heads, and then
compute $\hat{p}$ with it's (1-$\alpha$) confidence interval (CI)
utilizing the following approximations of the binomial parameter $p$
\cite{agresti1998approximate}:

\[
Wald's/Normal:\:\hat{p}=\frac{k}{n}\:\:\: CI:\:\hat{p}\pm z\sqrt{\frac{\hat{p}(1-\hat{p})}{n}},
\]

\[
Agresti-Coull:\:\hat{p}=\frac{k+\frac{1}{2}z^{2}}{n+z^{2}}\:\:\: CI:\:\hat{p}\pm z\sqrt{\frac{\hat{p}(1-\hat{p})}{n+z^{2}}},
\]

\[
Continuity\: Corrected\: Wilson\: Score\: Interval:\:\hat{p}=\frac{k+\frac{1}{2}z^{2}}{n+z^{2}}\:\:\: CI:\:(a,b),
\]

\[
where:\: a=max\left\{ 0,\frac{2k+z^{2}-\left(1+z\sqrt{z^{2}-\frac{1}{n}+4k(1-\frac{k}{n})+(4\frac{k}{n}-2)}\right)}{2\left(n+z^{2}\right)}\right\}, 
\]

\[
and\: b=min\left\{ 1,\frac{2k+z^{2}+\left(1+z\sqrt{z^{2}-\frac{1}{n}+4k(1-\frac{k}{n})-(4\frac{k}{n}-2)}\right)}{2(n+z^{2})}\right\}, 
\]

\noindent where $z=z{}_{\alpha/2}$, the value at which the standard normal
cumulative density function is equal to $1-\frac{\alpha}{2}$, and
$\alpha=$one minus the desired confidence level as a fraction.  Here, $z=1.96$
at a 95\%, (1-.05), confidence level, the confidence level used throughout the study. The interpretation of a confidence interval is fairly simple.
If given a CI, $\left(a,b\right)$, for $p$ at 95\% confidence where
$a$ and $b$ are determined by the observation of $K$, ideally 95\%
of the observations of $K$ will give $a$ and $b$ such that $\left(a,b\right)$contains
the true $p$ (for a general reference on approximations to binomial
proportions and confidence intervals see Ref \cite{devore2012probability}). This is not always
the case and actual coverage probabilities, the probability that a
CI formed from a random observation of $K$ will contain $p$, often
differs from the desired confidence level, especially at small $n$
and $p$. The amount this coverage probability varies from the desired
confidence level also depends on the confidence interval that is used
\cite{brown2001}.

Notice in the above equations that both the Wilson and Agresti-Coull confidence interval have
the the same $\hat{p}$, however the lengths and endpoints of the
confidence intervals differ. We perform the test on a range of $n$
values reflecting typical key term totals seen in the communities
recording $\hat{p}$ for each method and value of $n$. We repeat
the experiment 10,000 times. For each method, the sample standard
deviation of the $\hat{p}$ values as a percentage of $p$ was recorded
for every value of $n$ tested. This is an indicator of the reliability
in the estimates of $p$. As $\hat{p}$ is the same in both the Agresti-Coull
and Wilson's Interval, we expect this experiment to give nearly identical
results for the two, however the difference in the widths of their
confidence intervals will prove to be more important later in the
study. The same procedure is repeated for values of $n$ and $p$
reflecting typical values seen in the entire LCC of a graph.

Using the estimate we deem best suited for our study, for every key
term we were able to obtain an estimate of $\hat{p}_{comm}$ for every
community and $\hat{p}{}_{LCC}$ along with 95\% confidence intervals
for each $\hat{p}$%
\footnote{We realize that, excluding Wald's estimate, combining the $\hat{p}$'s
from all terms in a given group will not form a valid PMF for that
community/the LCC as they will not sum to one. However, if key terms
are viewed individually, Wald's interval is proven to be the poorest
confidence interval for $\hat{p}$, as the actual coverage probability
drops well below 95\%, especially for small values of $p$ and $n$.
Conversely, Agresti and Coull's and Wilson's intervals are shown to
have an actual coverage probability much closer to or above 95\%,
even at small $p$ and $n$ (all of our $\hat{p}$'s will be small
and, in several communities, so will $n$), while maintaining similar
confidence interval widths to Wald's \cite{brown2001}. Consequently,
we are getting a better estimate of $p$ without increasing the implied
uncertainty of the estimate $\hat{p}$ as a larger confidence interval
indicates more uncertainty in $\hat{p}$. This feature of Agresti
and Coull's and Wilson's intervals is ultimately essential for the
most pertinent use of these $\hat{p}$'s to this study - determining
which of a community's key terms are significant and represent the
topic of that community.%
}.

\subsection{Determining a Community's Significant Key Terms}

As stated before, for each $\hat{p}$ there exists an error associated
with the estimate related to the width of the (1-$\alpha$) confidence
interval, $W$.  Analogous to how the standard deviation, $\hat{\sigma}_{Wald}$,
in the estimator $\hat{p}_{Wald}$, is $\hat{\sigma}_{Wald}=\nicefrac{1}{2}(\frac{W_{Wald}}{z_{\alpha/2}})$,
we will consider the standard error, $\delta$, in our estimates $\hat{p}$
to be $\delta=\nicefrac{1}{2}(\frac{W}{z_{\alpha/2}})$. Thus, each
key term has a $\hat{p}{}_{LCC}$ with error $\delta{}_{LCC}$ and
in each community a $\hat{p}_{comm}$ with error $\delta_{comm}$.
To determine a community's significant key terms, find terms such
that: 
\[
\hat{p}_{comm}-\hat{p}_{LCC}>\beta\left(\sqrt{\delta{}_{comm}^{2}+\delta_{LCC}^{2}}\right)
\]
where $\sqrt{\delta{}_{comm}^{2}+\delta_{LCC}^{2}}$ attempts to capture
the uncertainty present in the difference $\hat{p}_{comm}-\hat{p}{}_{LCC}$
and $\beta$ is a constant that controls how many standard deviations
$\hat{p}_{comm}$ has to be above $\hat{p}{}_{LCC}$ for a term to be
considered significant. Choosing $\beta$ too small will result in
too many key terms being deemed significant and one still would not
have a clear idea of the topic of the community. Additionally, with
small $\beta$, it is much more likely for some of the returned key
terms to have true values of $p{}_{comm}$ not greater than $p{}_{LCC}$,
but due to estimation error, will have $\hat{p}_{comm}-\hat{p}_{LCC}$
above the minimum threshold. Choosing $\beta$ too large will likely
result in the wrongful identification of all key terms as insignificant.
We chose $\beta=2$ for this study such that a select few key terms
were returned and, for these key terms, $p{}_{comm}$ is almost certainly
higher than $p{}_{LCC}$ due to focus on that term in the community
and not because of errors in our estimates $\hat{p}_{comm}$ and $\hat{p}_{LCC}$.
The more significant key terms identified from this analysis, the
greater evidence of topic focus in that community.

\subsection{Determining a Minimum Threshold for Community Size}

Due to the increasing error in the estimation of $p$ as $n$ decreases,
performing this analysis becomes less likely to return significant
key terms as the number of total terms in the community decreases.
Therefore, it is expected that this analysis becomes far less effective
and useful below a certain community size.

Community size can be defined in several ways. From a more general
co-authorship network analysis viewpoint, it can be defined as the
number of authors or the number of papers in a community. More applicable
to this study is community size in terms of total key terms contained
in all of a community's papers. By plotting the number of significant
key terms versus community size for each of these community size metrics,
we attempted to identify a minimum threshold community size according
to each measurement for use of this analysis. This is intended to
serve as a guideline for which communities it would be prudent to
use this method to identify their topic.

\section{Results/Discussion}

\subsection{Choosing an Estimator for $p$, Finding a Range of $n$ where the
Estimate is Reliable}

\subsubsection{Choosing the Estimator}

\begin{figure}
\includegraphics[scale=0.4]{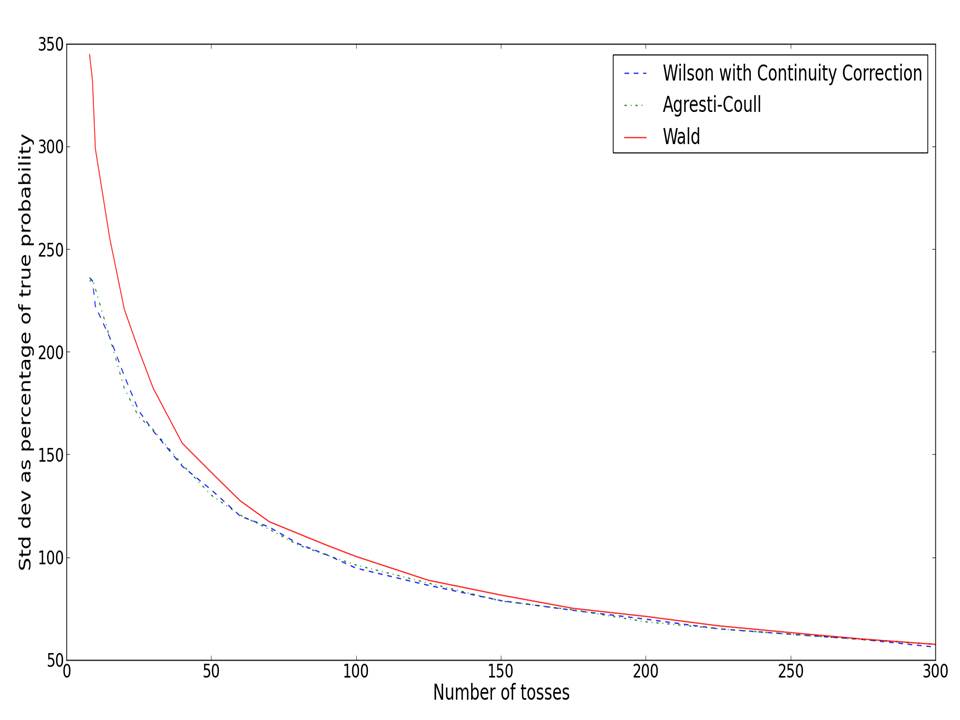}

\protect\caption{Sample standard deviation of 10,000 estimates of $p$ as a percentage
of $p$ versus number of tosses of a biased coin with probability
of heads appearing $p=0.01$. This is an indicator of reliability in
the estimator $\hat{p}$ for different values of $n$, the number of
Bernoulli trials performed, at true $p=0.01$, a value expected for
a borderline significant key term. The range of $n$ is chosen to
represent typical amounts of total terms present in smaller communities.
Three methods are used to calculate $\hat{p}$: Wilson's with Continuity
Correction (dashed), Agresti and Coull's (dotted), and Wald's (solid
line). The graph shows the Wilson and Agresti-Coull estimators have
less variability at small $p$ ($p=0.01)$ and small $n$, while all
three have very similar standard deviation at > 200 trials. This suggests
that either Wilson's or Agresti and Coull's estimate and confidence
interval should be used throughout the study. Additionally, it points
out that any estimate of $p$ below 150 terms has great uncertainty
as the standard error in the estimator is near or above 100\% of the
true $p$.\label{fig:CommFlips}}
\end{figure}
\begin{figure}
\includegraphics[scale=0.4]{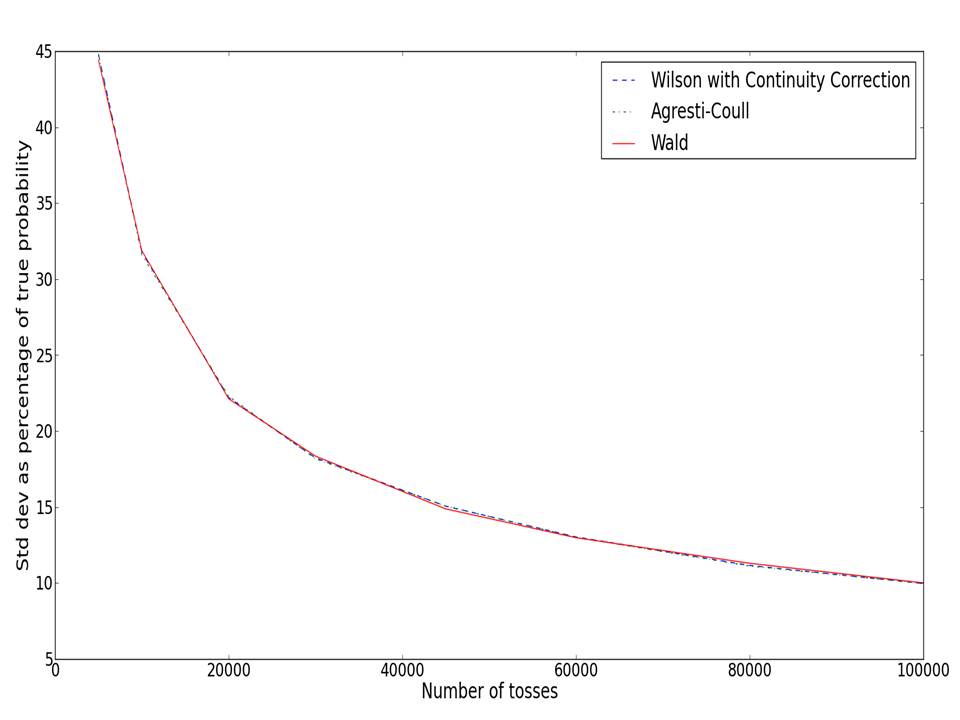}\protect\caption{Sample standard deviation of 10,000 estimates of $p$ as a percentage
of true $p$ versus number of tosses of a biased coin with $p=0.001$.
This determines the variability in the estimator $\hat{p}$ for different
values of $n$, the number of Bernoulli trials performed, at true
$p=0.001$, a value expected for terms across the entire LCC. The range
of $n$ is chosen to represent typical amounts of total terms present
in the LCC of a graph. Again, three methods are used to calculate
$\hat{p}$: Wilson's with Continuity Correction (dashed), Agresti
and Coull's (dotted), and Wald's (solid line). All estimates have
nearly the same standard deviations here, so any decisions for which
estimate to use for the procedure should be based off of considerations
for estimating $p{}_{comm}$, not $p{}_{LCC}$. Additionally, the
errors $\delta{}_{LCC}$'s are much smaller than $\delta{}_{comm}$'s.
For example, 25\% of 0.001 is much less than 100\% of 0.01. Thus when
finding the error in the difference of the estimates $\hat{p}_{comm}-\hat{p}{}_{LCC}$,
most will be determined by the error present in the estimator $\hat{p}_{comm}$.\label{fig:LCC Flips}}
\end{figure}

Figure \ref{fig:CommFlips} shows that estimating the true probability, $p$,
of success in a Bernoulli trial based on one observation of the count,
$k$, of successes in $n$ trials becomes less and less reliable as
$n$ decreases and the standard deviation of $\hat{p}$ increases
greatly. This is especially true for small $p$, which we will be
dealing with in a large majority of key terms. As mentioned before
and confirmed in Figure \ref{fig:CommFlips}, Wald's estimate of $p$
performs the worst for small $p$ and $n$ as there is great uncertainty
in the measurement. Wilson's and Agresti-Coull's perform better below
$n=200$, a common number of total terms for small communities, at
$p=0.01$, a typical value for a key term's $p{}_{comm}$, suggesting
one of these two should be used as the primary estimator for the procedure
described in Section III.C. Wilson's and Agresti and Coull's estimators
$\hat{p}$ are calculated the same way, so no difference is seen in the
variability of the estimator here. However, as stated before, the
widths, endpoints, and effectiveness of the 95\% confidence intervals
are the determining factor for which estimate and confidence interval
to choose for determining significant key terms. According to the
study of different binomial confidence intervals done by Brown, Cai,
and DasGupta, the Continuity Corrected Wilson confidence interval
has a shorter expected confidence interval width than Agresti and
Coull's while maintaining an actual coverage probability closer to
the desired 95\% than Agresti and Coull's especially in for small
$p$ and $n$ \cite{brown2001}. Therefore, based on the superior
performance of the Continuity Corrected Wilson confidence interval
in estimating $p{}_{comm}$, if it also performs at least as well as the
other two estimators in estimating $p{}_{LCC}$, we should find
it proper to use this estimator and confidence interval for the remainder
of the study. As a test of each estimate's performance in estimating
$p{}_{LCC}$, the three different estimates are evaluated again, but
now at $p=0.001$, a common $p{}_{LCC}$, and $n$ values representative
of typical values seen in a LCC. Figure \ref{fig:LCC Flips} shows
that all three estimators are equally variable for such values of
$n$ and $p$. It follows that any decisions for choosing an estimator
for the procedure should be based off of how the estimator performs
estimating $p{}_{comm}$, not $p{}_{LCC}$. Thus, this result confirms
our choice to use Wilson's confidence interval with Continuity Correction
as the estimator to implement the procedure described in Section III.C.

\subsubsection{Range of $n$ where $\hat{p}$ is Reliable}

Another noteworthy feature of Figure \ref{fig:CommFlips} is that,
even when using Wilson's estimate, the standard error in the estimator
will be great compared to the true value of $p$ - around or above
100\% of $p$ - when $n$ is less than 150. This suggests that our method
of comparing community and component PMF's to determine significant
key terms in a community, which relies on a moderately accurate
estimate of $p_{comm}$, may not be a reliable or effective procedure
for determining the topic of communities with less than around 150
total terms. Further analysis of 150 terms as a minimum community
size threshold takes place in Section IV.B.3.

Figure \ref{fig:LCC Flips} shows us that the errors in $\hat{p}_{LCC}$ will
be much less than the errors in $\hat{p}{}_{comm}$, both as a percentage
of true $p{}_{LCC}$ and in total error. This is because 10\% to 45\%
of 0.001 is much less than 50\% to 200\% of 0.01. Though the error in
$\hat{p}_{LCC}$ is present and affects the error in the difference
$\hat{p}_{comm}-\hat{p}{}_{LCC}$, it is almost negligible compared
to the error in $\hat{p}{}_{LCC}$. However, to err on the side of
caution, we still consider the standard error in the difference $\hat{p}_{comm}-\hat{p}{}_{LCC}$
to be $\sqrt{\delta{}_{comm}^{2}+\delta_{LCC}^{2}}$, the sum of the
errors in quadrature.

\subsection{Finding Significant Key Terms that Determine the Topic of a Community}

\subsubsection{A Look at How Significant Key Terms Represent the Topic of a Community}

Utilizing the Continuity Corrected Wilson confidence interval at 95\%
and the $\hat{p}$ that accompanies it, the procedure for finding
significant key terms in Section III.C was performed on networks of
various sizes. The following is an example of some results achieved
from using this method. In a network formed by searching the term
`neuromorphic computing' in Thompson Reuter's ISI Web of Knowledge
v.5.5 index according to Section III.A, 918 papers were returned to
form the network which also contained 1414 unique authors. This is
one of the smaller networks analyzed in this study. The LCC was formed
by 270 papers and 370 unique authors. One of the communities within
the LCC contained about 25 papers. Representative titles include:
`Analog VLSI models of range-tuned neurons in the bat echolocation
system', `Analog VLSI circuits for attention-based, visual tracking',
and `A Neuromorphic VLSI Head Direction Cell System Online Correction
of Orientation Estimates Using Spatial Memory in a Neuromorphic Head
Direction System'. Several significant key terms were returned for
this community. Among them included: `VLSI', `visual information',
`motion estimation', `mechanoreception', `bioacoustics', `biomimetics'.
Although the sample size is small, the qualitative match between the
significant key terms and the titles seems to accurately reflect the
themes of the papers. Therefore we argue it is acceptable to use these
terms to represent the topic of the community.

\subsubsection{Quantitative Measures for the Amount of Focus on a Topic}

As a quantitative measure of community focus, we could look at the
number of significant key terms in a community. If a community is
focused on a topic, we expect to see a lot of key terms related to
that topic repeated in several papers, causing a high $\hat{p}_{comm}$
for those terms and consequently being marked significant. Conversely,
if a community is unfocused, there will be plenty of key terms, but
few will be repeated in more than one paper. In this case, for most
terms $\hat{p}_{comm}$ will not differ much from $\hat{p}{}_{LCC}$.
When a community does have only a few significant terms it could be
the result of the following two characteristics of communities: 1)
either there is actually not much focus on any topic in the community,
or 2) the community was not large enough to obtain enough total terms
for a reliable estimate of $p{}_{comm}.$ If 1) is true, that will
result in small differences of $\hat{p}{}_{comm}-\hat{p}{}_{LCC}$
for nearly all terms, such that $\hat{p}{}_{comm}-\hat{p}{}_{LCC}$
does not exceed $2\sqrt{\delta{}_{comm}^{2}+\delta_{LCC}^{2}}$, our
minimum threshold for marking a key term significant. If 2) is true,
it causes a large error in $\hat{p}_{comm}$ as seen in Figure \ref{fig:CommFlips},
driving the error in $\hat{p}_{comm}-\hat{p}{}_{LCC}$ up enough that
even terms with moderately sizable differences in the estimates of
$p{}_{comm}$ and $p{}_{LCC}$, $\hat{p}_{comm}-\hat{p}_{LCC}$ will
not be greater than $2\left(\sqrt{\delta{}_{comm}^{2}+\delta_{LCC}^{2}}\right).$
In either case, we are not able to determine the community's topic,
if there is any clear topic, from our analysis.

The arguments presented in the previous paragraph
seem to suggest using the percentage of unique key terms in a community
that are significant as a measure of the amount of focus in a community. However,
we caution against using this measure when comparing communities with
a large difference in size. To see why, first consider a community
with only 50 unique terms. Having just 3 of them be marked significant
results in a 6\% percent rate. On the other hand, in a large community,
with a much larger sample size, say 1000 unique terms, having 50 significant
key terms results in a 5\% rate. It's not clear which is really more
focused: a community with a small sample size and only a few significant
key terms, but a higher significance rate; or a large community with
many more total terms and significant terms, but a lower success rate.
Therefore, we recommend only using this measure to compare communities
with similar size.

\subsubsection{Determining a Minimum Community Size Threshold for Using this Analysis}

\begin{figure}
\includegraphics[scale=0.35]{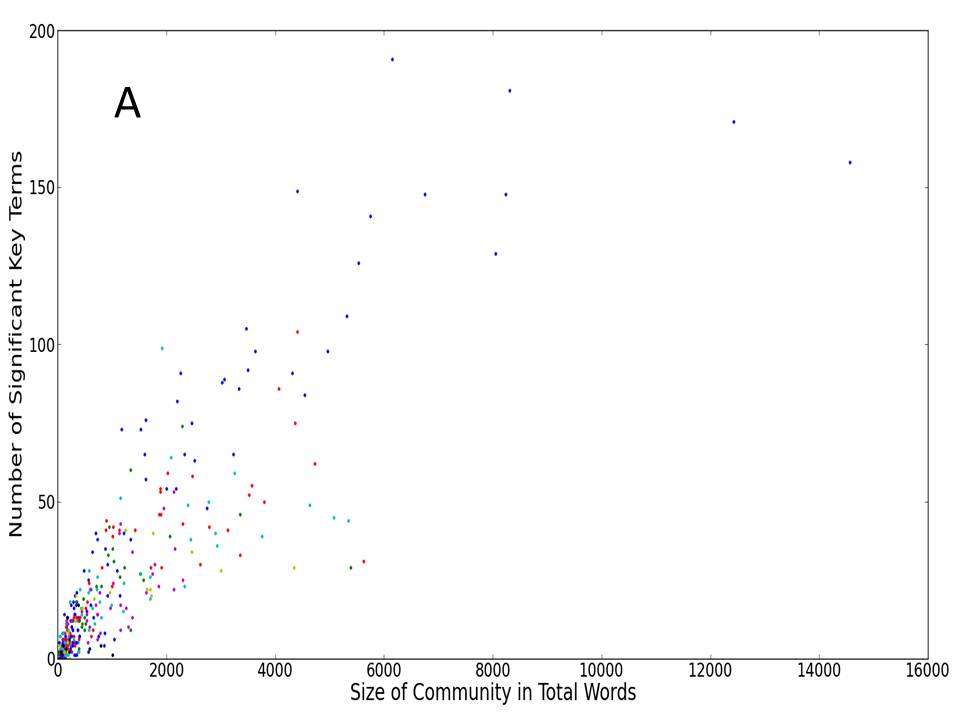}\includegraphics[scale=0.35]{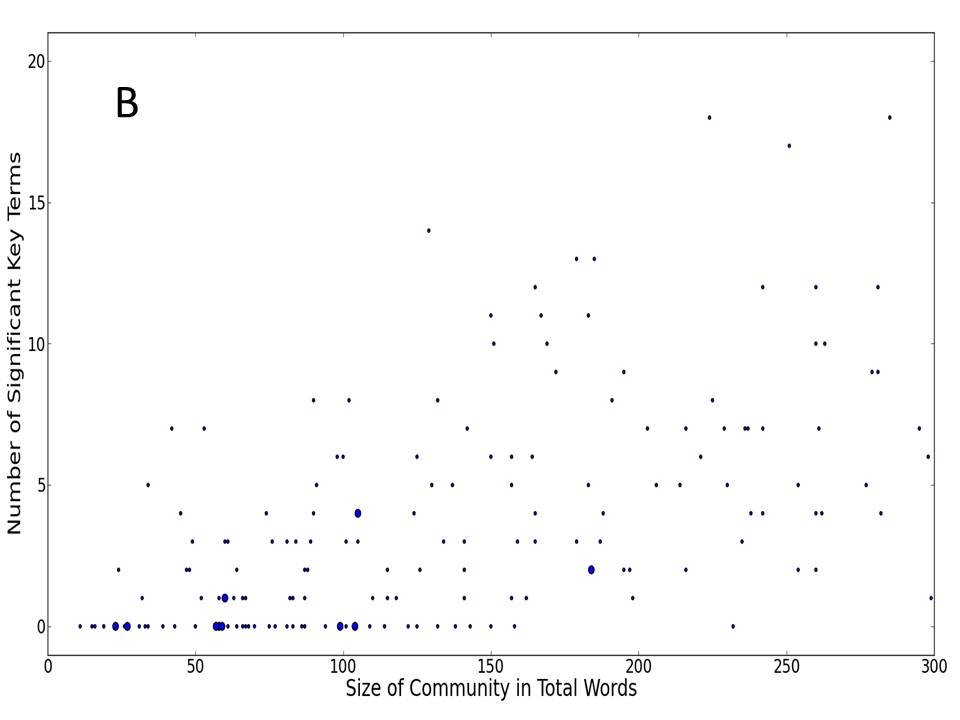}

\protect\caption{(Color Version Online) A. Number of significant terms per community
versus community size measured in total terms per community. Communities
from 8 different networks of various sizes ranging between 800 to
15000 publications are plotted here. It is clear that as the number
of total terms in the community increases, so does the number of significant
key terms. The fact that most communities contain many of these community-specific
significant key terms is an indicator that, in general, scientific
communities in co-authorship networks are focused on researching a
specific subtopic of their field. B. Similar plot to A, focused on
the densely populated region near the origin. The dot sizes indicate
how many communities share the same (x,y) data point. The larger dots
indicate there are 2 communities at that point while the smaller dots
indicate there is 1 community at that point. A noticeable change in
the graph occurs at 150 terms. Above this point, significant key terms
are almost always present in communities, and the number of these
significant key terms is occasionally high enough to make some conclusions
about the topic of the community. Below this point, often no significant
key terms are identified in the communities, and most of the time less
than 5 terms are marked significant. Our method to identify a community's
topic is therefore not effective in communities with less than 150
total terms.\label{fig:TotalWords}}
\end{figure}

Throughout the paper it has been suggested that this analysis may
not perform well in smaller communities where the total number of
terms is relatively low. In Section IV.A it was suggested that the
analysis would not be effective in communities with less than 150
terms. In an attempt to show that 150 is a reasonable lower limit
on the number of total terms needed to use this analysis on a community,
the number of significant key terms per community versus community
size in total terms was plotted in Figure \ref{fig:TotalWords} with
data from eight different networks.

In Figure \ref{fig:TotalWords}A, we see more significant key terms
are returned as community size increases. One reason larger communities
have more significant terms is that estimates of $p{}_{comm}$ become
less variable in communities with large $n{}_{comm}$. This causes
the error $\delta{}_{comm}$ to be smaller, and differences in $\hat{p}{}_{comm}-\hat{p}{}_{LCC}$
are more likely to be above $2\sqrt{\delta{}_{comm}^{2}+\delta_{LCC}^{2}}$.
Thus, our procedure is more effective in picking out significant terms
because we are getting more precise estimates of $p_{comm}$. A more
obvious reason for this is that there simply aren't as many unique
terms in small communities to choose from. Recall that the number
of total terms is not the same as the number of unique terms, but
they are closely related. Community size in total terms is equal to
$n_{comm}$, which is $\sum_{\left\{ terms\in\Omega\right\} }k_{comm}$,
the sum of all the counts for each term in $\Omega$ within the community,
whereas the number of unique terms in a community is the number of
the terms belonging to $\Omega$ such that $k{}_{comm}\neq0$. Communities
with larger $n{}_{comm}$'s tend to have more unique terms which leads
to more possibilities for significant key terms. However, this alone
is not a guarantee there will be any significant key terms. As discussed
earlier, a community, large or small, without a clear topic will have
little or no significant key terms. The fact that communities with
enough terms consistently return large numbers of significant terms
indicates many communities are focused on specific topics. Next, we
look at how the analysis performs in communities without enough terms,
and ascertain the minimum number of terms necessary for the analysis
to be effective.

As seen in Figure \ref{fig:CommFlips}, the variability of $\hat{p}_{comm}$ starts
to increase significantly below $n=150$. Thus, in Figure \ref{fig:TotalWords}B
we focus on the range of sizes near 150 terms to determine if this
is the right minimum threshold to use. At less than 150 terms, the
procedure for identifying significant key terms will return no significant
terms about half the time and less than 5 significant terms most of
the time. This is often not a large enough sample size of significant
key terms to deduce the topic of a community. So, whether the low
significant key term count is due to the lack of enough terms, or
the increasing error in $\hat{p}_{comm}$, or both, the procedure
is most often ineffective in communities with less than 150 total
terms. This agrees with the inference made about the effectiveness
of the procedure in communities with less than 150 terms made in Section
IV.A. In communities with size slightly above the threshold at about
150-300 terms, rarely does this method return no significant key terms,
and sometimes upwards of 10-15 terms are returned. Thus, the procedure
is worth trying in this range of $n$. Finally, as seen in Figure
\ref{fig:TotalWords}A, communities with greater than 300 terms almost
always return many significant key terms.

Though community size in total terms is very applicable to this study,
community size is rarely measured by the amount of total key terms
present in the `Topics' fields of the papers present in that community,
but rather, often measured by number of authors or papers. However,
this analysis allowed us to form a connection between the community
size and the number of significant key terms in a community. And,
using this relationship, we can extend the conclusions to find thresholds
for the more common measures of author number and paper count.

\begin{figure}
\includegraphics[scale=0.35]{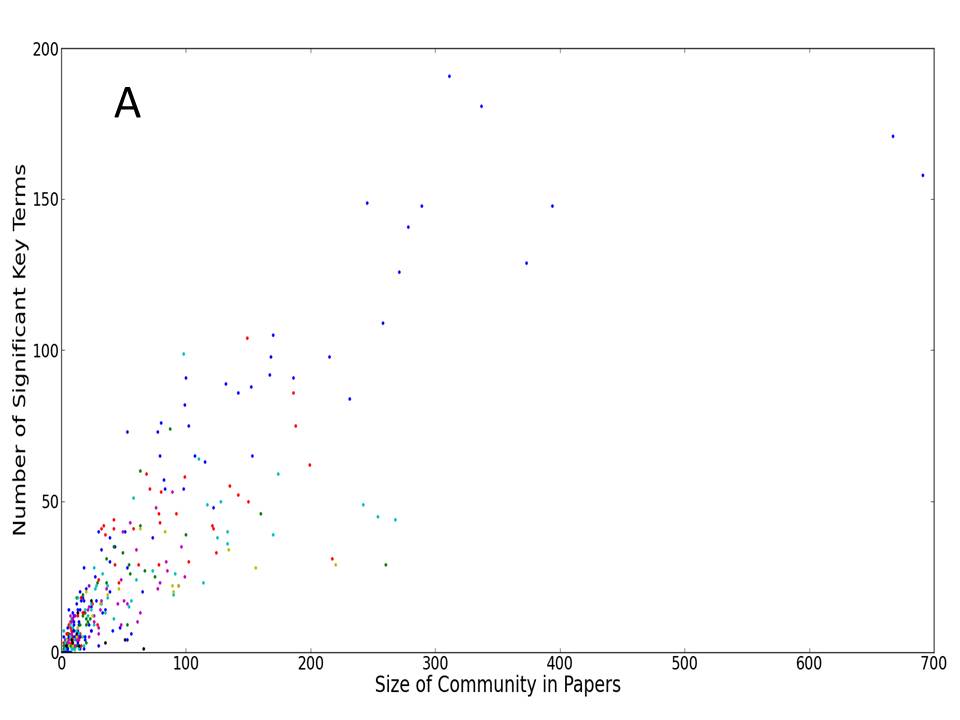}\includegraphics[scale=0.35]{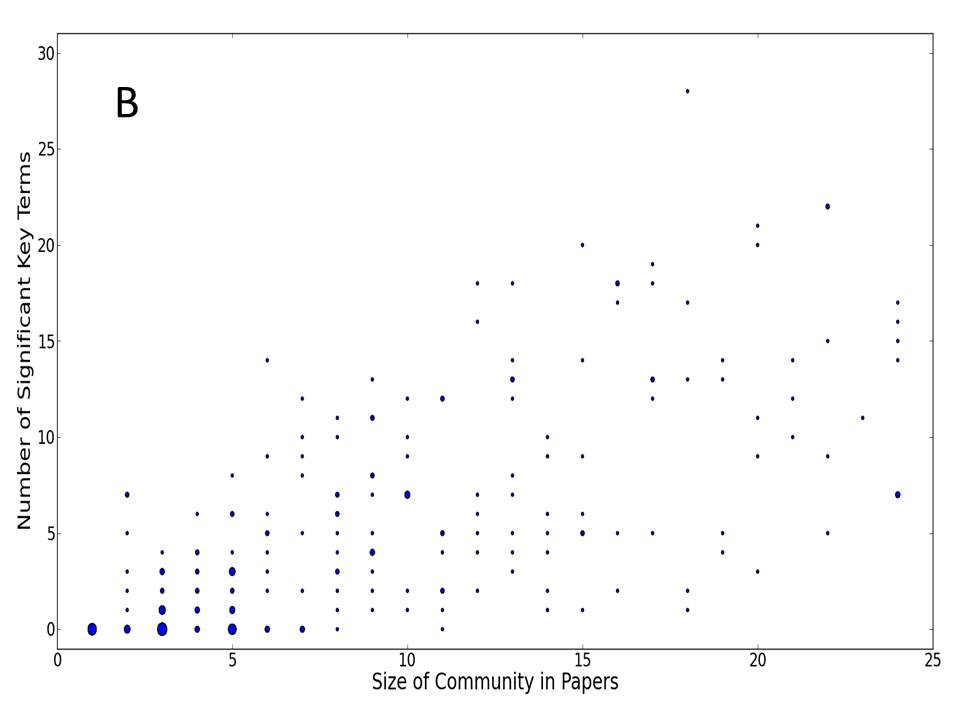}

\protect\caption{(Color Version Online) A. Number of significant terms per community
versus community size measured in number of papers per community.
Communities from 8 different networks of various sizes ranging between
800 to 15000 publications are plotted here. Each different color represents
communities from a different network. There is a nearly identical
relationship between number of significant key terms in a community
and community size in papers as seen in Figure \ref{fig:TotalWords}
which plotted significant key terms versus size in total terms. This
is because of the strong correlation between the number of papers
in a community and the number of total terms present. B. Similar plot
to A, focused on the densely populated region near the origin. The
dot sizes indicate how many communities share the same (x,y) data
point. The largest dot size represents 13 communities while
the smallest dots indicate 1 community at that point. Using
the same arguments to pick a minimum community size threshold for
recommended use of this procedure that were used to determine a threshold
for size measured in total terms, we suggest using this method in
communities with more than 8 papers.\label{fig:Papers}.}
\end{figure}

The relationship between the size of the community in papers and the
size in total terms is fairly straightforward. It is positively correlated,
as the number of papers increases, the number of key terms generally
increases also. Although, this is not always the case as there is
no set number of key terms per paper. So, for example, it is possible
for a community containing 20 papers to have more key terms than a
community with 25 papers if some of the authors in the 20-paper community
prefer to tag papers with many key terms. There is a very strong correlation,
0.992, between papers and terms though, so we expect plotting the number
of significant key terms versus community size in papers to reveal
similar patterns to Figure \ref{fig:TotalWords}. Figure \ref{fig:Papers}
confirms this is indeed the case.

Nearly all the same arguments used to choose a minimum size threshold
for number of terms apply here as well. The only difference is that
having less papers only strongly implies there will be less terms,
but it isn't a certainty. After that, the rest of the logic follows
as before. 8 papers appears to be the right threshold here. Additional
analysis of our data shows that a paper contains, on average, 21 key
terms. Since the correlation between community size in papers and
terms is very strong this indicates that a communities with 7 papers
would tend to have about 142 terms (below the term threshold) while
communities with 8 papers should have near 163 terms (above the term
threshold). This additional data supports the choice of 8 as a threshold
for community size in papers.

Finally, the relationship between the number of authors in a community
and the number of terms in a community is only slightly more complicated.
Firstly, the size of a community in authors is strongly positively
correlated with the size in papers. Although, a paper can have anywhere
from 1 to over 100 authors \cite{Newman2001_CoAuthorNet}, so, much
like the relationship between papers and terms, a community can have
less authors than another but more papers. Then, as described earlier,
the community size measured in papers is strongly correlated with
the size in total terms. As a result of this, the correlation between
the number of authors in a community and the number of terms- 0.917-
is less than the correlation between the number of papers in a community
and the number of terms- 0.992. This being said, we still expect many
of the same patterns to appear in the plot of the number of significant
key terms in a community versus the community size measured in authors
as in the previous two cases.

\begin{figure}
\includegraphics[scale=0.35]{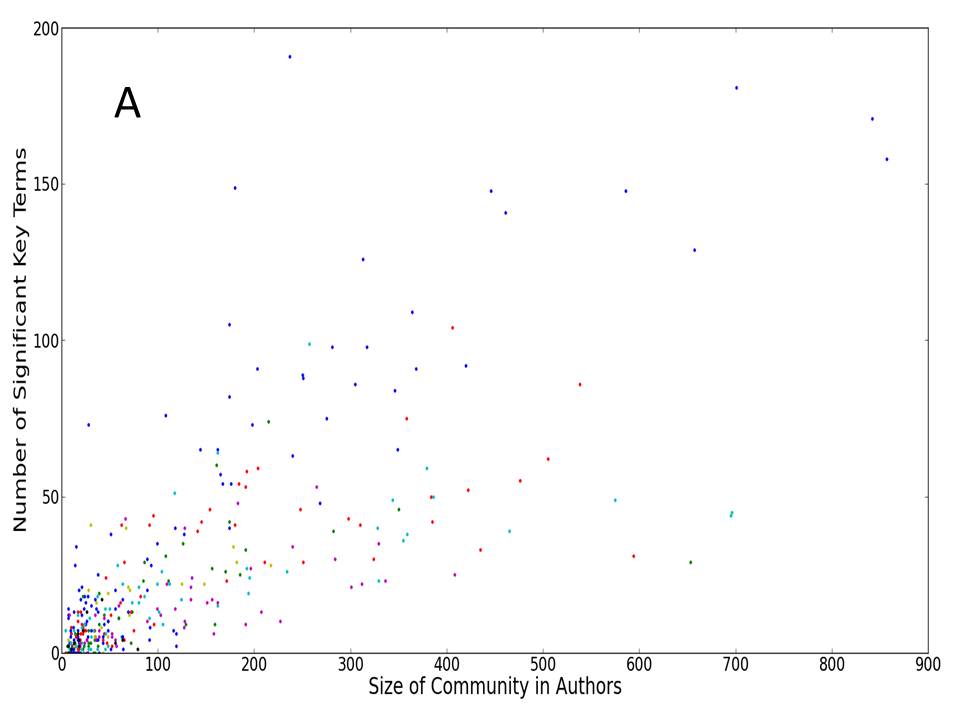}\includegraphics[scale=0.35]{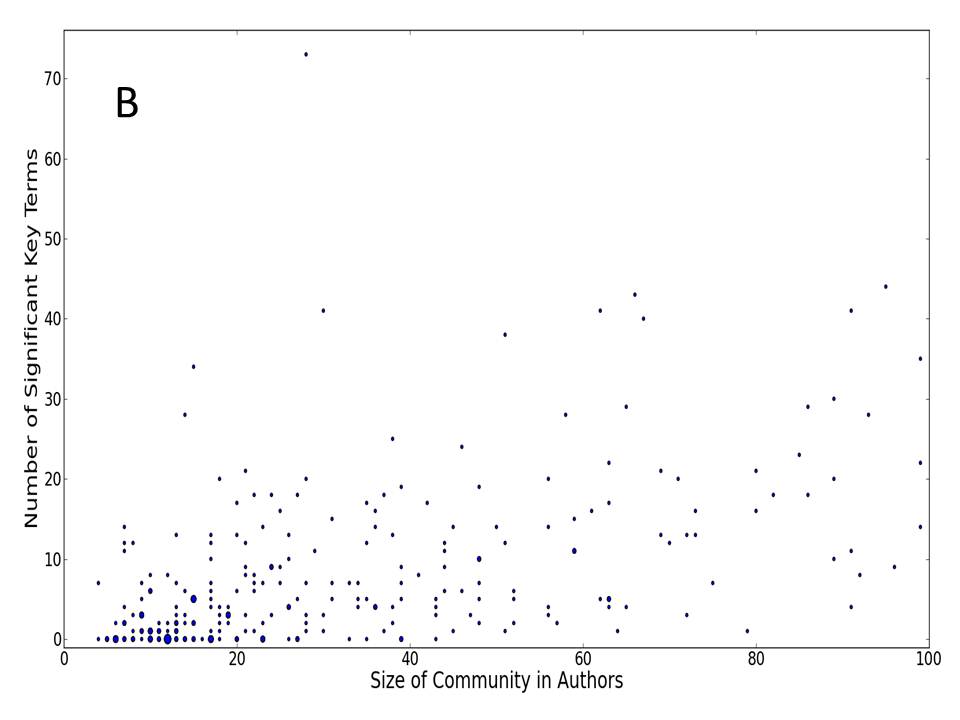}

\protect\caption{A. (Color Version Online) Number of significant terms per community
versus community size measured in number of authors per community.
Communities from 8 different networks of various sizes ranging between
800 to 15000 publications are plotted here. Each different color represents
communities from a different network. There is a very similar relationship
between number of significant key terms in a community and community
size in authors as seen in Figure \ref{fig:TotalWords} which plotted
significant key terms versus size in total terms. However, there is
less correlation between number of significant key terms and size
in authors than between number of significant key terms and size in
terms. B. Similar plot to A, focused on the densely populated region
near the origin. The dot sizes indicate how many communities share
the same (x,y) data point. The largest dot size represent 7 communities
at that point while the smallest dots indicate there is 1 community
at that point. Using the same arguments to pick a minimum community
size threshold for recommended use of this procedure that were used
to determine a threshold for size measured in total terms, we suggest
using this method in communities with more than 15 authors.\label{fig:Authors}}
\end{figure}

Figure \ref{fig:Authors} shows that many of the same patterns exist,
however the correlation between significant key terms in a community
and community size in authors isn't as strong as for the previous
two measures of community size. This is likely due to slightly decreased
correlation between the number of authors and the number of terms
in a community because of the more complex relationship between the
two. Using the same criteria for picking a threshold as before, we choose 15 authors as the minimum threshold for this
procedure to be somewhat effective (with effectiveness increasing
with size). Additional analysis of our data shows that, on average,
there are 10.5 key terms per author in a community. Since the correlation
between community size in papers and authors is fairly strong at 0.917
this indicates that a community with 15 authors would tend to have
slightly above 150 terms while communities with 14 authors are expected
to contain slightly below 150 terms. This additional data supports
the choice of 15 as a threshold for community size in authors. Though
if one was choosing between the two more common size measures- authors
and papers- we would suggest using community size in number of papers
and the threshold that corresponds to that size metric. This is because
there is greater clarity in how the number of papers affects the number
of total terms. Furthermore, by comparing Figures \ref{fig:Papers}
and \ref{fig:Authors}, one can see there are more communities
with author size above the 15 author threshold where the analysis
fails to produce significant key terms than with paper size above
its 8 paper threshold. Because the relationship between the number
of authors and total terms is more complicated (and less correlated),
it is difficult to tell if the absence of significant terms in these
communities is a result of lack of focus on a topic or because the
community is below the 150 term threshold (the strongest threshold
for this procedure). Conversely, if a community is above the paper
threshold, it is most likely above the 150 term threshold too. In
summary, the 8 paper threshold, while not as significant as the 150
term threshold, is more accurate than the 15 author threshold.

\section{Conclusion}

We have found an automatic method of using bibliographic data to determine
the research focus of communities present in co-authorship networks
that is most often successful in producing a representation of a community's
topic. By detecting key terms with a statistically significant greater
probability of appearing in a community than in the community's component,
we were able to find many significant key terms per community. Then,
through an inspection of a community's significant key terms, one
can deduce the community's topic.

Small communities are the exception where this method is not effective.
For a variety of reasons, few, or even no significant key terms are
returned. This is not necessarily an indicator that the community
lacks focus on a topic, but simply that there limitations in this
procedure which prevent us from being able to determine the topic
of small communities. Thus, a recommended community size threshold
is proposed for three different community size measures, above which
one will start to see some useful results from this method. For community
size measured in total terms: 150 terms; for community size measured
in papers: 8 papers; for community size measured in authors: 15 authors
(though we recommend using either of the first two community size
thresholds for analysis with this method). When these size requirements
are met and the method returns enough significant key terms, we have
seen qualitatively that examining those terms does, in fact, give
an accurate representation of the topic of the community.

Now, recall the question posed in the introduction: does the hierarchical
clustering present in the structure of the network imply a hierarchy
on the specificity of research topic also? The results produced from
the method to find significant key terms provide evidence that this
is the case. The method used was designed to pick out community specific
key terms: terms that appeared with great frequency in the community,
but not throughout the component. Most communities that were large
enough for this procedure to produce interpretable results contained
several of these community specific key terms. Since these terms represent
the topic of the community, it follows that the community has a specific
and unique topic. Thus, the division of the component into communities
is significant in the sense that it marks the division of a larger
group of scientists researching a more general topic common to the
entire component into small groups of researchers focused on researching
a specific subtopic of the field.

\section*{Acknowledgments}

The authors would like to thank Kathy-Anne Soderberg and Roberto Diener
for helpful discussions, and Mark Esposito for his support of this
project.

\bibliographystyle{unsrt}

\end{document}